\documentclass{article}

\usepackage{PRIMEarxiv}

\usepackage[utf8]{inputenc} 
\usepackage[T1]{fontenc}    
\usepackage{hyperref}       
\usepackage{url}            
\usepackage{booktabs}       
\usepackage{amsfonts}       
\usepackage{nicefrac}       
\usepackage{microtype}      
\usepackage{lipsum}
\usepackage{fancyhdr}       
\usepackage{graphicx}       
\graphicspath{{media/}}     
\usepackage{natbib}
\usepackage{graphicx}
\usepackage{placeins}
\graphicspath{ {./} }
\usepackage{multirow}
\usepackage{amsmath} 
\usepackage{float}
\usepackage{booktabs}
\usepackage{multirow}
\usepackage{enumitem}

\usepackage{booktabs}
\usepackage{longtable}
\usepackage{multirow}
\usepackage{threeparttablex}
\title{A rapid evaluation of Australia's COVID-era apprentice wage subsidy programs
}

\author{
  Peter Bowers \\
  Australian Department of the Treasury\thanks{This work does not necessarily reflect the position of the Australian Government or the Australian Treasury. Use of terms like `we' throughout refers only to the authors.} \\
  \And
  Patrick Rehill\\
  RMIT University\\
  Australian Department of the Treasury\footnotemark[1] \\
  \texttt{patrick.rehill@rmit.edu.au} \\
  \And
  Ethan Slaven \\
  Australian Department of the Treasury\footnotemark[1] \\
}

\begin{document}
\maketitle

\begin{abstract}
    In the midst of the COVID-19 pandemic in 2020, the Australian Government launched two programs to incentivise new apprentices to start and complete apprenticeships --- the Boosting Apprenticeship Commencements (BAC) and Completing Apprenticeship Commencements (CAC) programs. These programs were wage subsidies to encourage employers to take on or retain apprentices. This paper evaluates the impact of these programs on apprenticeship commencements and completions taking a mixed-methods approach combining econometric modelling and interviews with stakeholders including employers and peak bodies. The programs led to a 70\% increase in commencement of apprenticeships but do not seem to have boosted retention rates. Taking cancellation rates as an indirect measure of apprenticeship completions (as data was not available to measure completions directly), there appears to be a small increase in cancellation rates suggesting lower eventual completion rates compared to previous cohorts. Cancellation rates were higher for non-trade commencements (7\% increase) during BAC, but slightly lower for trade commencements (0.7\% decrease). We find this effect in non-trade apprenticeships was likely driven by `sharp practice' where some employers took advantage of the BAC by converting existing employees over to apprenticeships to attract the wage subsidy with no intention of having these employees stay as apprentices beyond the period of the BAC's generous subsidy. While the BAC / CAC were successful in many of their goals, the program was not perfect and there are several lessons that can be learnt from its design. In particular, the need to implement the program quickly meant early design choices inadvertently encouraged `sharp practice' and a rush for places that placed strain on the training sector. However, employers appreciated the front-loading of payments which provided the most financial support when apprentices were new and at their least productive.
\end{abstract}


\section{Introduction}
\begin{figure}
    \centering
    \includegraphics[width=0.75\linewidth]{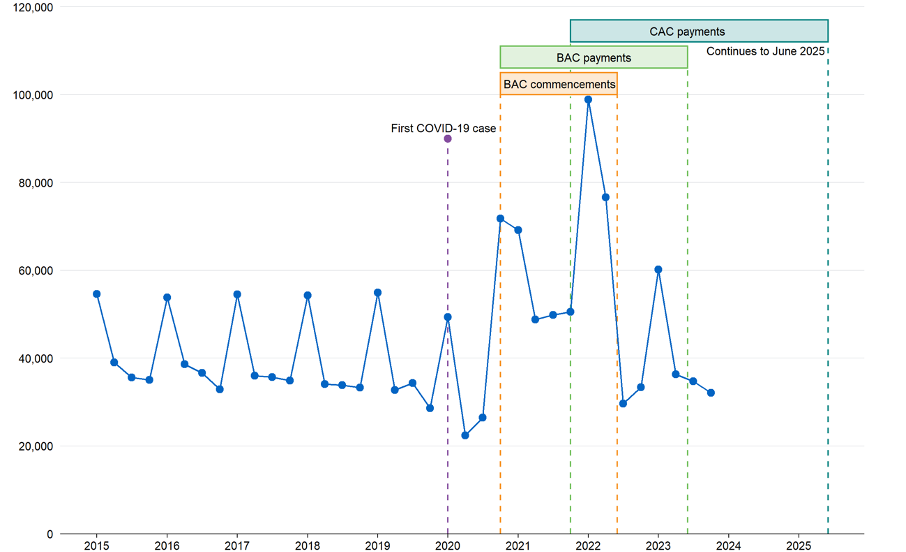}
    \caption{Key policy events and number of apprenticeship commencements by quarter (based on DEWR administrative data)}
    \label{fig:background}
\end{figure}

The Australian Government has long taken an interest in incentivising apprenticeships as a means of ensuring a pipeline of key workers, particularly in trade professions. This interest became particularly strong during the height of the COVID-19 pandemic in 2020 -- 2021 \citep{smith_expansion_2021, stanwick_issues_2021}. There was a concern that existing supports would not be enough to convince employers to continue taking on new apprentices or to retain existing apprentices. At this time there was substantial economic uncertainty and there were periods where containment policies --- such as travel and gathering restrictions --- made it difficult for some businesses to function as normal \citep{edwards_variation_2022}. As Figure \ref{fig:background} shows, in the first part of 2020, apprenticeship commencements were substantially down from previous years. A sustained fall off in the number of apprenticeships could have disrupted the pipeline of training meaning fewer opportunities for those leaving school in 2020 and 2021, leading to fewer skilled workers in the medium-term. Finally, a fall in employment of apprentices might have had a macro-level impact in deepening the general economic contraction during 2020. 

The Australian Government's response to this problem was a pair of wage subsidy programs. The first was the Boosting Apprenticeship Commencements (BAC) program announced on 4 October 2020, and the second was the Completing Apprenticeship Commencements (CAC) program announced in December 2021 as part of the 2021-22 Mid-Year Economic and Fiscal Outlook. These were essentially wage subsidy programs designed to encourage commencements of apprenticeships (BAC) and retention of apprentices (CAC). These programs were designed to incentivise `Australian Apprenticeships', which include both traditional apprenticeships (focused on trades such as building, plumbing and carpentry), and also apprenticeships and traineeships of other kinds (such as retail and hospitality). For convenience, we refer to all apprenticeships and traineeships as `apprenticeships' throughout this paper and distinguish between trade apprenticeships and non-trade apprenticeships based on the occupation classification of each role.\footnote{Trade occupations are defined as occupations in Australian and New Zealand Standard Classification of Occupations (ANZSCO) Major Group 3 --- Technicians and Trades Workers}

The BAC program provided a wage subsidy for the first year of an apprenticeship, covering 50 per cent of its gross wage for the 12-month period from the date of commencement; up to a maximum of \$7,000 per eligible apprentice per quarter (which equates to \$28,000 for the 12-month period).

The CAC program provided a tapered wage subsidy over the second and third year of the apprenticeship (only for those eligible for BAC in the first year). This covered 10 per cent of an apprentice’s gross wage for the period from 12 to 24 months from the date of commencement; up to a maximum of \$1,500 per quarter (which equates to \$6,000 for the 12-month period). It then covered 5 per cent of an apprentice’s gross wage for the period from 24 to 36 months from the date of commencement; up to a maximum of \$750 per quarter (which equates to \$3,000 for the 12-month period). Ultimately, the programs were open to people who commenced as an apprentice between 5 October 2020 and 30 June 2022 before these programs were ended. Details on these programs were provided in the Department of Employment and Workplace Relations’ (DEWR) Australian Apprenticeships Incentives Program Guidelines \citep{department_of_employment_and_workplace_relations_australian_2023}. 

\subsection{This evaluation}

The focus of this evaluation was on the effect of the BAC and the CAC programs on commencement and completion of apprenticeships. It aimed to understand if the programs boosted commencements and completions and if so, for whom. This evaluation uses an event study approach to model commencements and uses a probit model fit on historical cancellations to study cancellation of apprenticeships. To be clear, while `event study' can mean many different things, we mean it in the sense used by \citet{huntington-klein_effect_2021}. We discuss heterogeneity in these results across different characteristics of employers and apprentices. Finally, we use interviews with stakeholders as part of an explanatory mixed-methods approach \citep{plano_clark_mixed_2022} to try and explain the drivers that led to the results. This analysis helps to focus on which aspects of the program worked, which didn't, and where the program was not used as intended. This paper is based on the Australian Centre for Evaluation's evaluation report \citep{australian_centre_for_evaluation_rapid_2025} which contributed to a \textit{Strategic Review of the Australian Apprenticeships Incentive System} released in January 2025, independently led by Dr Iain Ross AO and Ms Lisa Paul AO PSM.

The evaluation finds that the BAC program substantially increased commencements (by 70\%) but the effect on expected completions was expected to be smaller. While the evaluation was undertaken too early to measure actual completions, we could indirectly measure likely completions by measuring early cancellations. The programs had little measurable effect on completion rates for trade apprenticeships (cancellation rates decreased by just 0.7\%) while non-trade apprenticeship cancellation rates rose by 7\% during the program. Interviews suggest this was likely due to `sharp practice' where some employers used the system in unintended ways, shifting regular employees into apprenticeships to attract a wage subsidy. This is not to say that the programs together did not lead to an \textit{absolute} increase in completions, rather the increase in commencements was accompanied by a decrease in the completion rate for non-trade apprenticeships.

\section{Literature review}

The Australian apprenticeships system is characterised by combining on-the-job training as an employee, government supports, and off-the-job training through vocational education and training (VET) providers (these are specialised training providers, generally not high schools as is the norm in many other countries) \citep{smith_expansion_2021, international_labour_organization_incentives_nodate}. Broadly there are three main approaches used around the world to incentivise employers to take on apprentices --- tax incentives, a levy scheme which pays some of an apprentices wage from a fund that all employers in a sector must pay into, and direct subsidies to employers. \citep{international_labour_organization_incentives_nodate,cie_financial_2024}. The Australian system relies largely on the latter. Under the subsidy system, government directly pays employers to take on apprentices across several different programs. This means they can take a more active hand in targeting particular sectors \citep{international_labour_organization_incentives_nodate}. It is important to note as well that governments have an important role in funding VET education in secondary and tertiary education (for example Australian Technical and Further Education institutions commonly known as TAFEs), but this is not directly relevant to this evaluation \citep{stanwick_issues_2021, owen_giving_2016}.

In order to understand the role of government around apprenticeships, it is important to understand that Australia is --- and was at the time studied in this paper --- experiencing a historic decline in the number of apprenticeships. This has led to skills shortages in occupations where training is delivered through apprenticeships \citep{productivity_national_2020}. This problem was particularly severe during the early part of the COVID-19 pandemic. Not only were commencements down over time, as shown in Figure \ref{fig:background}, but suspensions --- that is an employer or apprentice suspending an apprenticeship --- had increased roughly four-fold on numbers in 2019 \citep{stanwick_issues_2021}. This was the environment in which the government decided to introduce the BAC and CAC programs.

The policy settings around apprenticeships in Australia are complex and often-changing. This is complicated by the fact that responsibility for different parts of apprenticeship policy are split across federal and state / territory governments. At the time the BAC and CAC were introduced, Australian governments already had incentives and supports for apprentices in place including wage subsidies or payments to employers, but none of these were as general or as generous as the two programs studied here \citep{deloitte_access_economics_australian_2020}.
\footnote{Note that this literature relies largely on government reports to provide background on apprenticeships in Australia as there is little peer-reviewed literature on this specific subject in Australia \citep{stanwick_issues_2021}. In addition, many of these reports are unpublished and have been shared with the Australian Centre of Evaluation by DEWR for the purposes of this evaluation. There are however several secondary sources that summarise this work that are public though including \citet{deloitte_access_economics_apprentice_2021}, \citet{deloitte_access_economics_australian_2020}, and \citet{anao_design_2024}.}

In fact, the BAC and CAC amounted to a ten-fold increase on spending to directly support apprenticeships \citep{australian_centre_for_evaluation_rapid_2025}. However, there were many smaller or more targetted supports already in place. A 2022 summary of Australian Government programs lists 22 different payments to subsidise apprenticeships, most of which are quite targeted \citep{department_of_employment_and_workplace_relations_australian_2023}. This is not to count additional state and territory schemes that also exist to support apprenticeships.

There have been several previous evaluations of incentive programs for apprenticeships in Australia. Rather than enumerating all of these, this review will focus just on those where full evaluation reports were available to the research team. It is worth briefly noting that the supports government offers to encourage commencing and completing apprenticeships are much broader than just incentive programs including professional development programs for those supervising apprentices, mentorship programs and advisory services for apprentices \citep{stanwick_issues_2021, owen_giving_2016}. However, for the purposes of this work, we are interested just in the incentives aspect of this set of policies. In addition, while there is a reasonably long history of support programs, we will focus here largely on recent programs --- i.e. those from the past two decades.\footnote{For a more expansive literature review see \citet{stanwick_issues_2021}.}

The Australian Apprenticeship Incentives Program (AAIP) commenced in 1998 and made payments to employers of any apprentice to encourage commencement and completion. These payments were typically \$1500 at commencement and \$2500 at completion. An evaluation from \citet{deloitte_access_economics_australian_2020} found that under this program commencements declined over the decade (from a peak of 330,000 in 2010-11 to around 155,000 in 2018-19), and argues much of this effect was due to the withdrawal of incentives for existing workers in occupations not on the National Skills Needs List in 2012-13. The drivers of commencements over 2010-2020 appeared to be specific to individual occupations and specific state and territory economic and policy changes. Trade-based apprenticeship commencements had not grown with population and employment growth as they were largely stable over 2013/14-2018/19. The implicit training subsidies arising from the AAIP were generally higher for non-trade-based traineeships than trade-based apprenticeships because non-trade-based traineeships were of a much shorter duration. Evidence of the effect of incentives on completion rates was mixed \citep{deloitte_access_economics_australian_2020}.

The Australian Apprenticeship Wage Subsidy (AAWS) was a program trialled in 2019 aimed at regional employers taking on apprentices in occupations aligned to the National Skills Needs List (NSNL). This was evaluated by \citet{deloitte_access_economics_australian_2022} but this evaluation did not use counterfactual methods, instead it simply examined the change in time-series after the introduction. The AAWS provided a wage subsidy at 75\% of the award wage in the first year, 50\% in the second year and 25\% for the third year (and no subsidy in the fourth year). Overall, \$120 million in wage subsidies were allocated and the program supported 2,715 new apprenticeships. These 2,715 rural and regional apprenticeship commencements represented 13\% of all rural and regional apprenticeship commencements in skills need occupations in 2019. Total rural and regional apprentice commencements increased by 1.3\% (523) over 2019. From 2015 to 2018, the number of commencements fluctuated around 41,000 but overall declined by 0.3\% on average year-on-year. Total participation by rural and regional employers increased by 2.0\% (760) over 2019, after average year-on-year declines of 0.6\% from 2015 to 2018.

The Supporting Apprentices and Trainees (SAT) scheme was like the BAC and CAC, a COVID-era support program which applied to existing apprentices employed between January 2020 and March 2021. It was also evaluated using non-counterfactual methods \citep{deloitte_access_economics_supporting_2022}. It provided a  wage subsidy that was valued at 50\% of apprentice wages to a maximum of \$7,000 per quarter per apprentice. It supported small and medium sized employers to retain existing apprentices and trainees, and re-hire apprentices displaced during the pandemic. It dwarfed the previous two programs with \$1.9 billion in wage subsidies provided to nearly 75,000 employers supporting over 150,000 apprentices. This was not a targeted incentive, but a broad program designed to keep apprentices employed during a time of economic uncertainty and in many places `stay-at-home orders' to restrict movement and transmission of COVID-19. Over the course of the program, the number of employers participating in apprenticeships increased over 2020 (by 4\%) and 2021 (by 16\%), after year-on-year declines in the previous five years. The total apprentice volumes and continuing apprentice volumes increased over 2020 (by 4\%) and 2021 (by 20\%).

Existing evidence suggests that employers differ substantially in how responsive they are to apprenticeship incentives --- and more responsive firms may also have worse completion outcomes. \citet{bednarz_understanding_2014} argues that the most subsidy-sensitive employers tend to have lower completion rates, drawing on apprentice interviews in \citet{dickie_fair_2011}. This work describes highly responsive employers as generally smaller, less established, with weaker ties to industry networks, and often only taking a single apprentice at a time. Conversely, employers who regularly engage apprentices, have better ties into the wider industry and who are often larger are better placed to support apprentices through to completion. Evidence from Germany shows a similar pattern with larger firms having higher training quality on average with better job outcomes for apprentices after they finish \citep{muller_size_2018}.

However, the evidence does not point one way here. A pair of evaluations by Deloitte Access Economics \citet{deloitte_access_economics_econometric_2012,deloitte_access_economics_apprentice_2021} show that employers with a longer history of hiring apprentices were more responsive to financial incentives. One possible explanation for this is that these two pieces of research just have different scopes; the Deloitte Access Economics reports focus on all occupations, while Dickie et al. focus just on trades. Another framing which might explain these differences comes from Powers \citet{powers_bricklaying_2013} who conducted a cluster analysis of employers of bricklaying apprentices in Australia across sensitivity to subsidies and completion rates. This paper found that while employers with higher sensitivity to subsidies had a lower completion rate overall there was one cluster of employers (27\% of those studied) that had both a high completion rate (over 90\%) and high sensitivity to subsidies. While it is a study of one trade, it does highlight that there may be a lot of heterogeneity in the kinds of firms that are responsive to incentives.

Another useful lens for distinguishing between sensitive and insensitive employers is the reason they engage apprentices. \citet{pfeifer_firms_2016} discusses two models of motivation to hire apprentices in Australia --- the `production model' and the `investment model'. In the former case, apprentices are brought on for the primary purpose of being low-wage workers. Their value is in the short-term, substituting for regular labour while being paid less than a regular employee. In the latter case, apprentices are primarily a longer-term human capital investment. Eventually they will be able to do high-value skilled work tailored to the employer's needs. Investment model apprenticeships are generally more valuable to governments looking to target apprenticeship incentives because they more deliberately build human capital \citep{pfeifer_firms_2016}, and because former apprentices who completed these kinds of programs tend to have better employment outcomes \citep{schuss_value_2021,cabus_productivity_2021}.

To the extent that incentives can boost commencements, there is some evidence that production model employers are more sensitive to financial incentives to engage additional apprentices because decision-making around hiring is much more tied to a short-term cost-benefit calculus \citep{pfeifer_firms_2016,schuss_value_2021}. In particular, a front-loaded subsidy like the BAC / CAC may have a particularly large divide here as such a program does reduce labour costs but does little to speed up the development of human capital. It is worth remembering though Powers' research which suggests (at least in the field he studied) that there can still be employers that are both sensitive to incentives and which intend to invest in supporting and developing their apprentices.

This production model versus investment model framing may tie into differences between trade and non-trade occupations in Australia. \citet{pfeifer_firms_2016} argues that apprenticeships in trades in Australia are generally more investment-oriented with non-trade occupations being much more production oriented. Non-trades generally do not have licensing requirements making an apprenticeship necessary and do not have long histories of formal apprenticeships / traineeships. Non-trade commencements were also much more sensitive to the removal of apprenticeship commencement payments for occupations not on the National Skills Needs List in 2012 \citep{mcdowell_shared_2011,pfeifer_firms_2016}.

In general, it seems like incentives can have a real effect on decisions around apprenticeships, particularly for targeted programs \citep{deloitte_access_economics_australian_2022, deloitte_access_economics_supporting_2022, deloitte_access_economics_australian_2020, laundy_apprenticeships_2016,cie_financial_2024}. However, the quality of this evidence is mixed. Few evaluations of these incentives have been carried out \citep{stanwick_issues_2021,cie_financial_2024}. Of those that have, there are few that have used experimental or quasi-experimental counterfactual-based methods.

More evidence exists on effects of similar programs internationally. Financial incentives to employers have been shown to be broadly effective in increasing commencements, though there is no clear evidence that they increase completions \citep{cie_financial_2024}. The two most relevant studies from overseas are from Czechia \citep{hora_empirical_2018} which ran an active labour-market program designed to place young people with limited work experience in wage-subsidised traineeships, and in Germany \citep{bonin_effects_2013} where there was a subsidy for employers engaging disadvantaged school-leavers. The Czech study did find evidence that the program reduced unemployment among young people, although there was some evidence of a substitution effect with subsidised trainees displacing regular employees (suggesting a production model approach by employers --- understandable given the program was primarily an active labour market policy for those at risk of unemployment, not one focused primarily on achieving vocational education goals). The German program is the more directly relevant, however it found the program had little effect on commencements and that the program typically did not affect hiring decisions at all. The payment was then best seen as just a windfall for employers who happened to already have hired eligible young people.

There is then not much robust published evidence that exists about the effectiveness of programs like the BAC and CAC, at least within the Australian apprenticeships system. This study aims to contribute evidence on the effectiveness of wage subsidies and to help build out some of the theoretical framework for understanding these policies with its qualitative component as well.

\section{Methods}
This section describes two kinds of model: the commencements model and the completions (cancellations) model.

\subsection{Commencements model} \label{sec:methods_com_mod}
\subsubsection{Data}
Commencement data is sourced from Apprentice and Traineeship administrative data provided by DEWR. This is population-level data on every Australian Apprenticeship. The raw data is contained in three separate tables that can be linked by `apprenticeship ID': the Activity Ledger table, Training Contract table, and the Employer table. The commencement series is taken from the Activity Ledger table at the apprenticeship level, where a commencing apprenticeship is the first observed entry of an apprenticeship ID. The occupation (ANZSCO unit) and location (state) of the apprenticeship is added from the Training Contract table, where occupation is linked to the apprenticeship qualification. The employing business’ industry (ANZSIC class) code is sourced from the Employer table. These variables (occupation, location, industry) are included as controls up until Q4 of 2019.

We also include commencements modelling results at the state and occupation level. These are calculated by running the model above on data that is specific to the state (for state results), or specific to the occupation (for occupation-level results). The time-varying economic and labour market control variables are: 

\begin{itemize}
    \item Hours actually worked by occupation (ANZSCO unit) and state are sourced from Table EQ08, ABS LFS Detailed (ABS, 2024).
    \item Unemployment rates by occupation (ANZSCO major) and state are sourced from Table UQB3, ABS LFS Detailed (ABS, 2024).
    \item Vacancy levels by occupation (ANZSCO unit) and state are sourced from Jobs and Skills Australia’s Internet Vacancy Index (IVI), Table ANZSCO4 Occupations, States and Territories (JSA, 2024). 
    \item JSA’s Internet Vacancy Index (IVI), largely made up of SEEK job ad data, is used as a measure of labour demand. The series contains job ad data by ANZSCO unit (4-digit) level and state which is aggregated up to the relevant occupation and geography level used in each model.
    \item Business confidence by industry (ANZSIC division) is sourced from the NAB Monthly Business Survey which is provided to Treasury (NAB, 2024).
\end{itemize}

Series were joined by occupation, industry and location. Data was then aggregated to higher-level occupation and location series. In this aggregation, count variables such as commencements and job ads are summed, and rate and index variables are averaged based on the number of commencements in each group (i.e. a mean of the aggregated series weighted by commencements in the disaggregated sub-series).

\subsubsection{Methods}

In order to model the effect of the BAC on commencements it is necessary to have a counterfactual with which to compare observations. This is made difficult because the policy was rolled out for the whole country at once making traditional approaches like difference-in-differences or synthetic controls infeasible. The analysis instead used a predictive model to estimate counterfactual outcomes and take the difference between observed and counterfactual outcomes as a causal effect. This is called an event study by \citet{huntington-klein_effect_2021}, or may also be called a statistical process control. This is not an event study in the way the term is generally used in econometrics which is to say a dynamic difference-in-differences. It relies on a stationary time series pre-treatment conditional on controls. Figure \ref{fig:bac_effect} shows that the time series is essentially stable with seasonality in the previous four years. It also shows that early divergence from this trend in COVID-19 before the introduction of the BAC / CAC is modelled well with the controls chosen. The full model is as follows.

For each quarter q at event time t ($t=0$ at BAC introduction) $$y_{t} = \alpha + \sum_{j=-4}^{10} \delta_j I[t = j] + \beta T_{tq} + \sum_{i=1}^{4} \gamma_i I[q = i] + \omega \mathbf{X}_{tq} + \varepsilon_{tq}
$$
where:
$y_t$ is log apprenticeship commencements at time $t$.
$\delta_j$ are the coefficients of interest that give the effect of a quarter being during BAC (or just before or just after BAC), on commencements. Note that $I[t=j]$  is a set of BAC variables for the 4 quarters before BAC, 7 quarters during BAC, and 4 quarters after BAC.
$T_tq$ is a vector of time variables, including a linear time trend with a break at Quarter 2 2012 to reflect structural changes in non-trade commencements due to a policy change. Time dummy variables are also added for Quarter 2 2012, Quarter 2 2013 and Quarter 3 2013 to reflect sudden and temporary spikes in non-trade commencements.
$I[q=i]$  is a set of variables to control for quarterly seasonal effects.
$X_{tq}$ is a set of time-varying economic and labour market control variables detailed immediately below.

A non-parametric bootstrap was used to calculate confidence intervals. This involves running the analysis many times with different datasets generated from the original by sampling with replacement. This can be used to empirically construct confidence intervals.

The choice of controls here is based on a literature review conducted by DEWR which is included in Appendix A.

\subsection{Completions (cancellations) model}

\subsubsection{Measuring completions}
The completions analysis used the same data sources as the commencements analysis. The only major change in data was in the outcome. One difficulty with analysing completions is that few BAC commencements could be expected to have had enough time to record a completion. For apprenticeships in trade occupations, most complete within 3-4 years, and in non-trades within 1-2 years. However, the administrative data on apprenticeships only goes to 1.5 years after the end of the period where commencements were eligible for BAC (BAC closed to new entrants June 2022, and the administrative data made available to the research team only goes to December 2023). Further, an apprenticeship may be suspended for a period and resumed later, and there is typically a time lag between a suspension, completion, or cancellation event occurring, and the employer reporting that event and it appearing in the data.

To get around these issues, this analysis constructs an indirect measure of likely completions by grouping each BAC apprenticeship into quarterly ‘commencing cohorts’, and categorises them as having either ‘progressed’ or ‘cancelled’ at the 18-month mark. For example, there could be 15,000 apprenticeships in the Quarter 1 2014 commencing cohort, 9,000 of which progressed after 6 quarters and the remainder – 6,000 – cancelled.

\subsubsection{Model}
For this modelling we use a probit model. Similarly to the modelling of commencements, this model is used to estimate a counterfactual for comparison.

The cancellation of an individual apprenticeship, $c_i$, is described with a probit model:$$c_i = \Phi \left( \alpha + \delta \text{BAC}_t \times \text{trade}_o + \sum_{j=1}^{4} \gamma_j \mathbb{I}[q = j] + \omega X_i + \beta E_{t,o,s} + \varepsilon_i \right).$$
For commencing cohort $t$ in occupation $o$ and state $s$:
$c_i$ is the probability of cancellation for an individual
$BAC_t$: indicates whether the apprenticeship was commenced during the BAC period.
$trade_o$:  indicates whether the apprenticeship was in a trade or non-trade occupation.
$X_i$: a set of time-invariant demographic variables: sex, age (and age squared), Indigenous status, born in Australia, English spoken at home, and self-assessed disability status; as well as apprentice-level industry fixed effects.
$E_{t,o,s}$:  are time-variant labour market variables, including unemployment rate by major occupation and state and job vacancy rate by detailed occupation level and state. These variables are entered both at the time of commencement and 4 quarters later to allow for economic conditions later in the apprenticeship to affect cancellation behaviour.
$I[q=i]$  is a set of variables to control for quarterly seasonal effects in cancellation rates.

\subsection{Interpreting effects as causal}
It is worth briefly discussing the causal interpretation of both models. 

With the event study analysis for commencements, the key assumption for identification is that the time-series is stationary conditional on control variables. The ideal approach would be to use a design with an observed control group allowing for the estimation of an average treatment effect on the treated (using a method like synthetic control or difference-in-differences). Unfortunately, as the program was rolled out to the whole country at once, this was not possible. Instead we take an approach that bases counterfactual outcomes during the period on pre-program outcomes. 

\begin{figure}[H]
    \centering
    \includegraphics[width=0.75\linewidth]{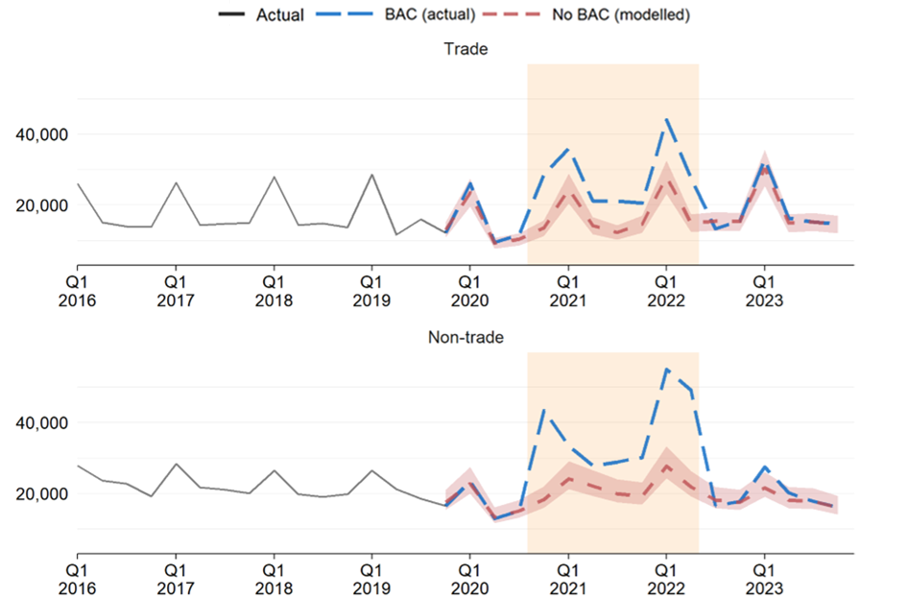}
    \caption{Comparison of commencements and forecasted commencements without BAC by quarter.}
    \label{fig:bac_effect}
\end{figure}

This event study amounts to a kind of lagged outcomes unconfoundedness design relying on an ignorability assumption \citep{huntington-klein_effect_2021, imbens_causal_2015}. The crux of the identifying assumptions then is whether the controls used to predict outcomes can make accurate predictions during the COVID-19 period. Essentially, we assume the effect of the pandemic is moderated through the controls chosen. This is a difficult assumption to validate. However, there are two different tests that can provide an indicator of how well this approach identifies the causal effect. The first is to essentially test counterfactual fit out-of-sample from Q4 2019 to Q3 2020. As shown on Figure \ref{fig:bac_effect}, these periods are pre-BAC / CAC but after what might be called the `training period'. The counterfactual line fits this data well, diverging only once the treatment begins. This is a good sign for identification.

The second test involves checking whether there are substantial COVID effects that are not controlled through the control variables. If there were substantial bias we would expect to see very different effects estimated for jurisdictions that were hit hard by the pandemic compared to those that were not hit hard. Results for Western Australia discussed in the next section show there does not seem to be such an effect. Importantly of course, even Western Australia was not spared the effects of the pandemic (with the state border being closed through most of the period), but the effect was relatively low compared to other states and territories, particularly those in the South-East of the country.

In the absence of randomisation or the possibility of a credible quasi-experimental design, this design represents a pragmatic middle-ground between a model with robust causal identification through an experiment or a convincing quasi-experiment and the approaches that past evaluations of similar programs have taken as discussed in the previous section.

These same assumptions must hold for the cancellations analysis as well. However, as this is an analysis of the rate of cancellations rather than the absolute number, these assumptions must only hold for the proportion of individuals cancelling, not the absolute number which is driven up by the commencements effect. Figure \ref{fig:cancellations-time} shows that there is good fit in the time series before treatment begins, although in the two quarters immediately before the start of the BAC, the model somewhat overestimates cancellations. If this overestimation continued, it would suggest a slight downwards bias in the treatment effect estimates.

\subsection{Interviews explaining results}

Interviews were used to help build out a theoretical framework for the quantitative findings. This amounts to a kind of explanatory mixed-methods approach where deductive quantitative analysis is followed by inductive qualitative analysis to explain findings \citep{plano_clark_mixed_2022}. We conducted seven interviews with stakeholders. These stakeholders included Group Training Organisations and one of their peak bodies, Australian Apprenticeship Support Services providers and their peak body, state regulators and TAFE Directors Australia.

\FloatBarrier
\section{Results}

\subsection{Commencements}
\subsubsection{Average level effects}
Over the seven quarters it was in effect, the BAC program resulted in a 70\% increase in commencements compared to the modelled counterfactual of what would have occurred in the absence of the program (Figures 2 and 3). This represents 191,000 additional commencements in total across the period studied (with a 95\% confidence interval of 154,000 to 222,000). The BAC program resulted in a significant increase in commencements in all quarters, but particularly in the first and final quarters of the program. For the first quarter, this was likely due to the program initially having a funding cap of 100,000 places, which resulted in employers rushing to commence new apprenticeships before all places were used up. For the final quarters, this was likely due to a “pull-forward effect” where employers sought to commence apprentices before the program closed to new entrants.

The effect was particularly strong for non-trades, where there were 114,000 additional commencements attributable to the BAC. This was 75\% higher than what would have been expected in the absence of the program (95\% confidence interval of 94,000 to 131,000).  For trades there were 77,000 additional commencements attributable to the BAC which was a 63\% increase compared to what would have been expected in the absence of the program (95\% confidence interval of 60,000 to 92,000).

For trades, commencements increased by around 100\% for the first and final quarters of BAC, compared to the number of commencements expected without BAC. Between these quarters, commencements were lower but still around 50\% higher than modelling suggests would be expected without the BAC. For non-trades, the effect was even stronger. In the first and final quarters of the BAC, non-trade commencements were almost 150\% higher and 135\% higher, respectively, than would have been expected without BAC. Between these periods, quarterly effects ranged from 30\% to 100\%.

 \begin{figure}[H]
     \centering
     \includegraphics[width=0.75\linewidth]{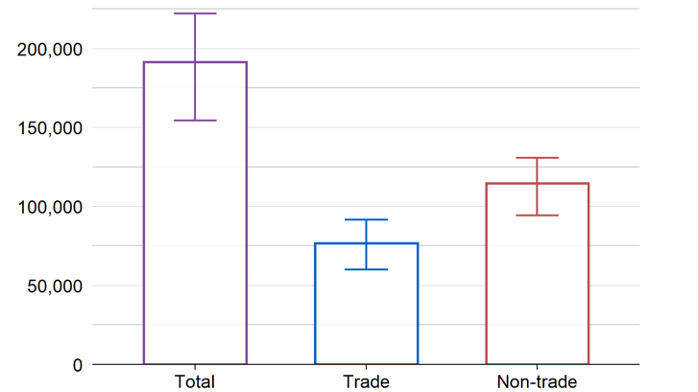}
     \caption{Effect of BAC on the number of commencements by trade and non-trade occupations.}
     \label{fig:bac_effect_het}
 \end{figure}
 
\begin{figure}
    \centering
    \includegraphics[width=0.75\linewidth]{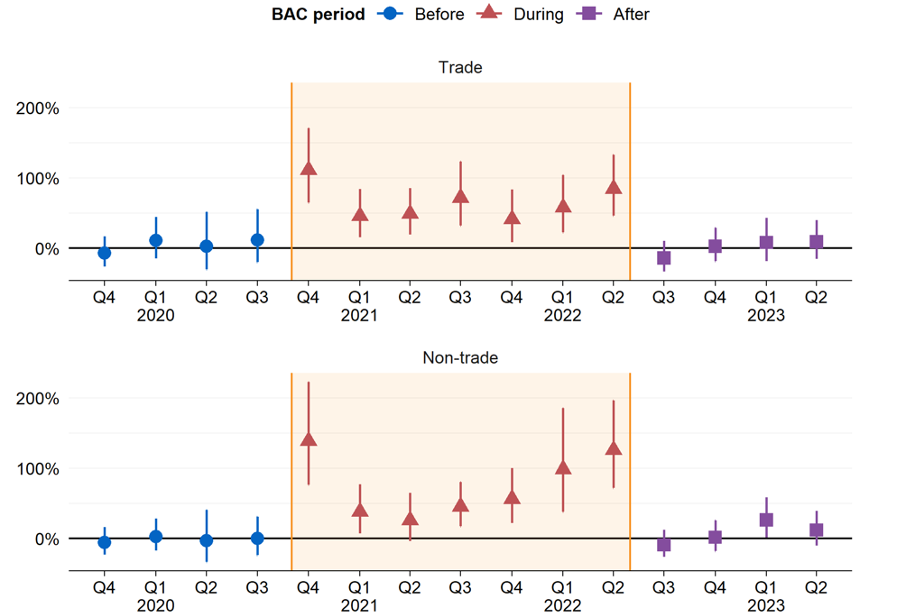}
    \caption{BAC quarterly effects on apprenticeship commencements, with seasonal and economic controls (error bars here represent a 99\% confidence interval for the effects)}
    \label{fig:bac_quarter_effects}
\end{figure}

\begin{figure}[!htb]
    \centering
    \includegraphics[width=0.75\linewidth]{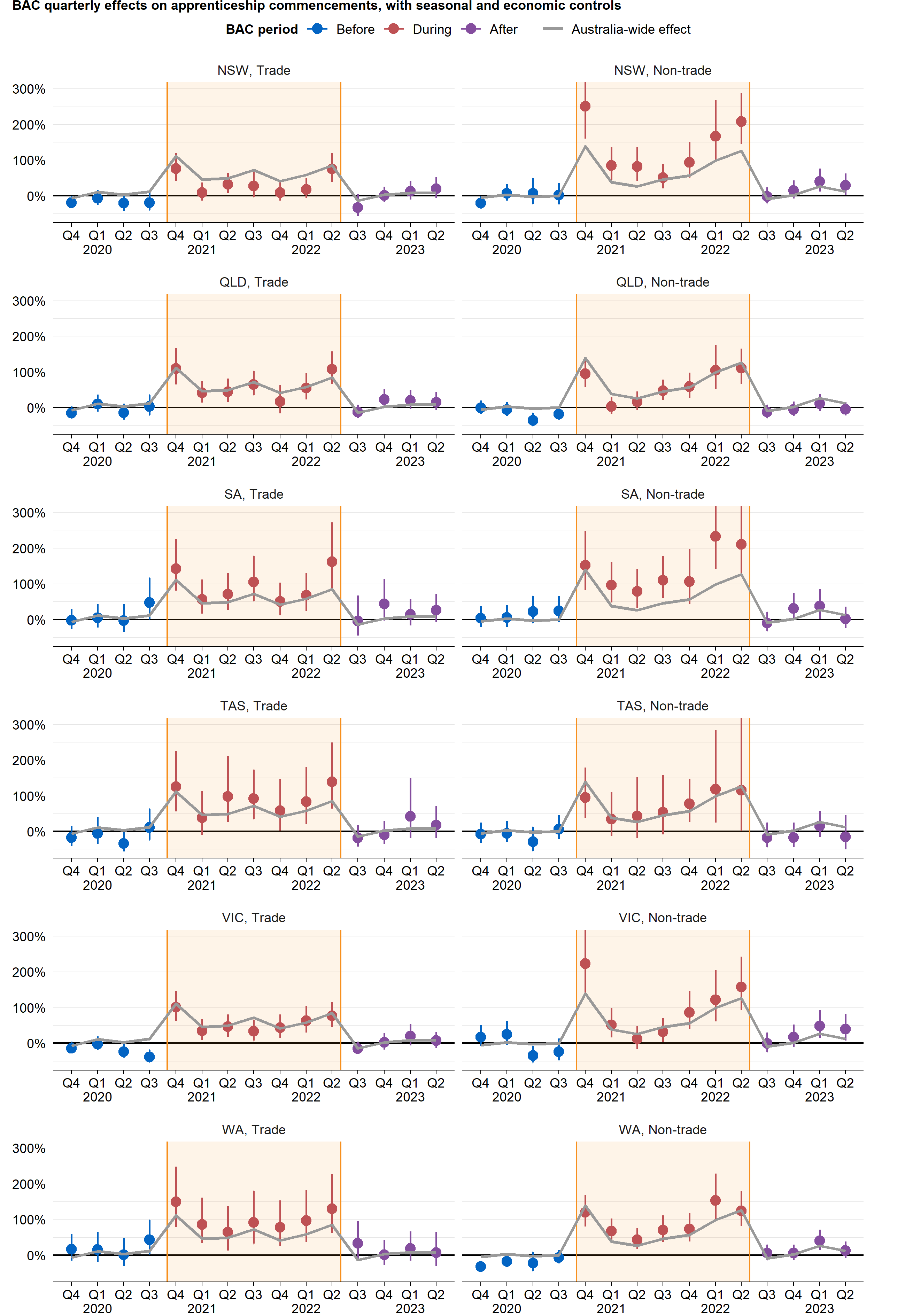}
    \caption{BAC quarterly effects on apprenticeship commencements by state / territory, with seasonal and economic controls (error bars here represent a 99\% confidence interval for the effects)}
    \label{fig:state_commence}
\end{figure}

There is some variation by state / territory as shown in Figure \ref{fig:state_commence}. There are two useful results here: a comparison that can be drawn between more and less affected states and territories; and specific impacts in states most affected by the pandemic-related lockdowns and COVID transmission. Firstly, in this design it is difficult to control for the effect of the COVID-19 pandemic if there are effects that are not mediated through the controls chosen because there is no control group that does not experience the pandemic. However, there is a quasi-control in the form of Western Australia which due to its geographic isolation avoided many of the most direct impacts of the pandemic (lockdowns or significant spread of disease \citep{edwards_variation_2022}). Therefore we can take the Western Australia results as a kind of quasi-control that shows an effect with minimal impact from the pandemic. Interestingly, the Western Australian results are not substantially different from those in other jurisdictions.

Another interesting pattern here is the substantial jump in commencements in the first quarter of the program for Victoria and New South Wales. These are the two jurisdictions for which we have results here that were hit hardest by the pandemic (NB: the ACT also experienced some significant lockdowns in 2021, however data on the territory was suppressed due to the small population size)\citep{edwards_variation_2022}. Employers may be more enthusiastic to take places given the higher economic uncertainty in the states that arose from strict containment policies. The fact these peaks are so much higher in non-trades occupations suggests possible sharp practice as discussed in the next section.

\FloatBarrier

\subsubsection{Heterogeneity across industry}

The effect of the BAC program on commencements was fairly consistent across trade occupations, however it varied more widely for non-trades (Figure 18). Occupations were analysed using the 97 Australia and New Zealand Standard Classification of Occupations (ANZSCO) Minor Groups. For example, ‘Automotive Electricians and Mechanics’ is an ANZSCO Minor Group that belongs to the ‘Technicians and Trades Workers’ ANZSCO Major Group. The following results should be treated with some caution given the smaller sample sizes for individual occupations which may mean substantial sampling error.

Commencement effects were consistently positive across all trade occupations: 
\begin{itemize}
    \item The three largest trade occupations – carpenters, electricians and automotive electricians – saw average commencements over the BAC period of around 40\% more than would have been expected without BAC. Hairdressers’ commencements increased by 39\% and wood trades workers had similar effects with a 53\% increase.
    \item By comparison, plumbers (increase of 30\%), mechanical engineers (increase of 25\%), and panel beaters (increase of 34\%) had less pronounced though still positive effects.
    \item Food trades workers – which includes chefs, cooks, bakers and butchers – had commencements that were 60\% higher than would have been expected without the BAC.
\end{itemize}

Effects for non-trade occupations were even greater than trades, with some occupations having particularly extreme growth in commencements. It is worth highlighting a few specific results.
\begin{itemize}
    \item Three of the four non-trade apprenticeships with the largest number of commencements all had strong increases due to the BAC program: general clerks (150\% increase), salespersons (123\% increase), and hospitality workers (110\% increase).
    \item Child carers had a smaller effect (12\% increase) that was not statistically significant. Discussions with stakeholders suggested this may have been due to changes in licensing requirements prior to the BAC period, already elevating these apprenticeship numbers to high levels. 
    \item Office managers had commencement growth far exceeding other occupations, with commencements 588\% higher during the BAC period than would have been expected without the program.
\end{itemize}

Some non-trade occupations were significantly lower than the counterfactual without the BAC program. We do not interpret this as the BAC program ‘backfiring’ by discouraging new commencements. Instead, this likely reflects other factors influencing commencements that were not captured in our model. For instance, personal service and travel workers occupations were greatly affected by border closures during the BAC period and saw a significant decrease of 39\% in commencements despite the BAC. Many travel agents had to find alternative employment, so employers were not taking new apprentices. 

\begin{figure}
    \centering
    \includegraphics[width=1\linewidth]{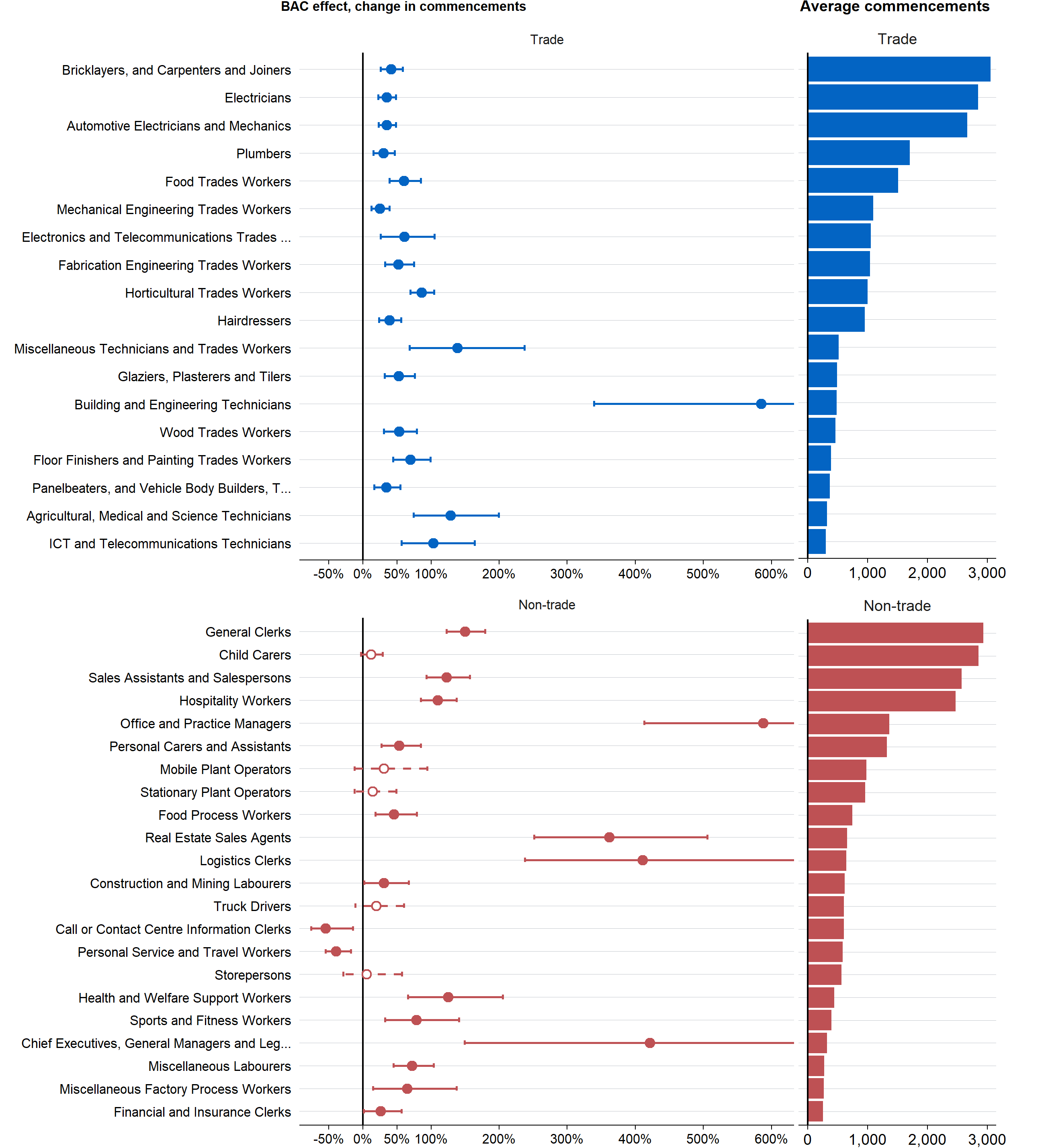}
    \caption{BAC's effect on trade and non-trade occupation commencement rates. Note: the average commencements is the mean across the BAC period. Results not significant at $\alpha=0.01$ are shown with hollow points and dashed error bars.}
    \label{fig:bac_by_industry}
\end{figure}

\FloatBarrier
\subsection{Completions (cancellations)}
\subsubsection{Average level effects}

\begin{figure}[!h]
    \centering
    \includegraphics[width=0.75\linewidth]{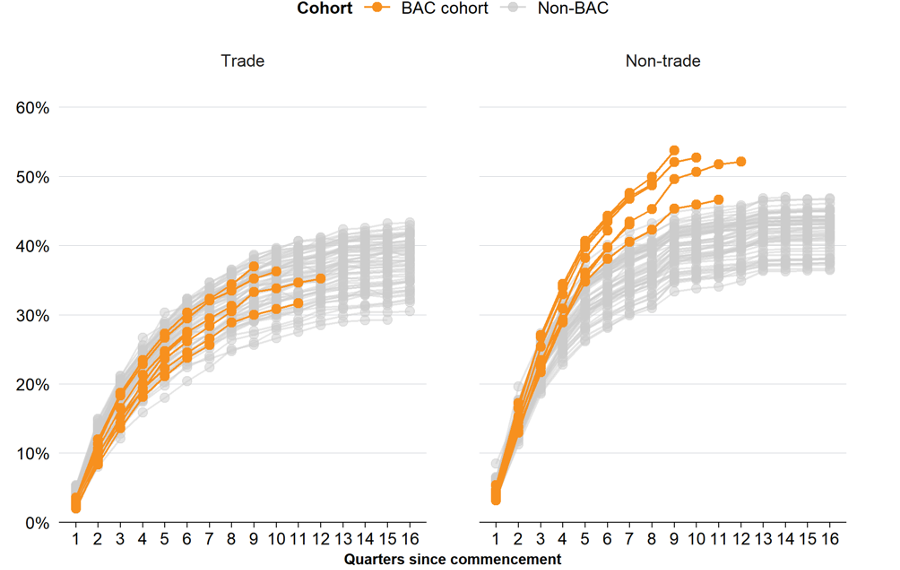}
    \caption{Cancellation rate by commencement cohort}
    \label{fig:cancel_cohort}
\end{figure}

BAC cancellation rates were 7\% higher than their historical levels for non-trades, but 0.7\% lower for trades. For non-trades, at 6 quarters after commencement, cancellation rates for all-but-one of the BAC cohorts were higher than previous cohorts (Figure \ref{fig:cancel_cohort}). For trade commencements, the cancellation rates for BAC cohorts were within the historical range, with between 25-30 per cent of apprenticeships cancelled 6 quarters (18 months) after commencement.

\begin{figure}
    \centering
    \includegraphics[width=0.75\linewidth]{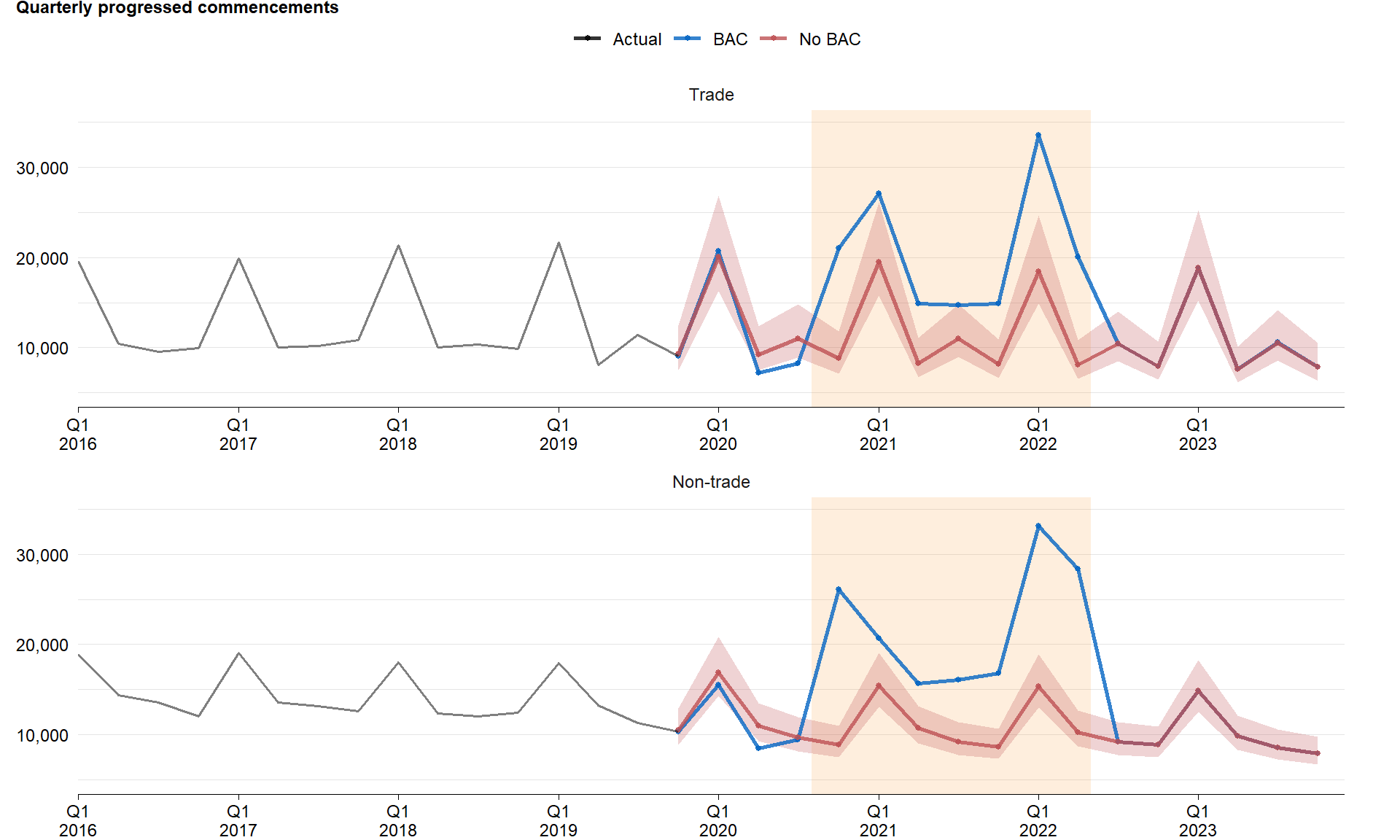}
    \caption{Observed and counterfactual outcomes across time for cancellations.}
    \label{fig:cancellations-time}
\end{figure}

Figure \ref{fig:cancellations-time} shows that the there are substantial spikes for both trade and non-trade occupations. However, in trade occupations, this spike is driven more by an increase in apprentices in absolute terms, which in the non-trade case there is also a change in the rate of cancellation as shown in Figure \ref{fig:cancel-het}.

\FloatBarrier

\subsubsection{Effect heterogeneity}

\begin{figure}[h!]
    \centering
    \includegraphics[width=1\linewidth]{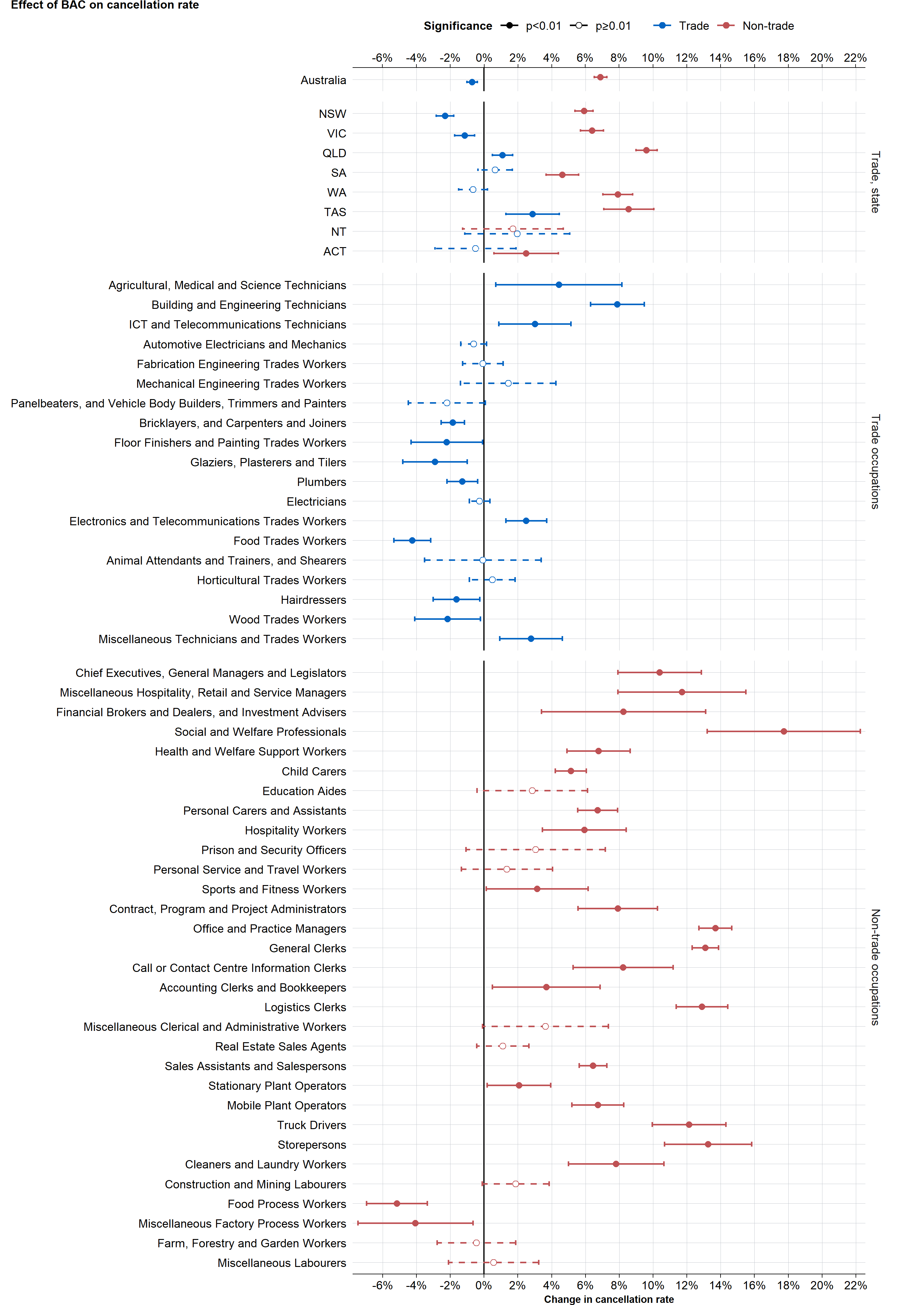}
    \caption{The effect of the programs on cancellation rate by jurisdiction and occupation. Subgroups with statistically significant results at $\alpha = 0.01$ are shown opaque while results than are not statistically significant are shown as semi-transparent (greyed out).}
    \label{fig:cancel-het}
\end{figure}

At the national level, the cancellation rate for trades was slightly lower, while for non-trades it was substantially higher. This is consistent with the descriptive statistics shown in the previous sections but has the advantage of controlling for compositional changes in commencements, and changes in labour market conditions.
Trade commencements during the BAC program were 0.7 percentage points less likely to be cancelled within the first 18 months. Non-trade BAC commencements, however, were about 7 percentage points more likely to be cancelled. This is a material shift when compared to the previous 6-quarter cancellation rate of 30-40 per cent for non-trades. These results were consistent across all states and territories (Figure 7).
Occupations with higher BAC commencements tended to also have higher cancellation rates (Figure \ref{fig:cancel-het}). In particular, the following occupations all had high commencements during BAC, and also had some of the highest cancellation rates: Office and Practice Managers; Chief Executive, General Managers and Legislators\footnote{The reason for such a large cohort of apprentices in this category is unclear. They are likely senior managers undertaking something like a Cert IV in Leadership. This may or may not be in-line with the intent of the program and could reflect some amount of sharp practice (the high cancellation rate in Figure \ref{fig:cancel-het} could be interpreted as suggesting this). However, this kind of traineeship is an entirely legitimate Australian Apprenticeship.}; General Clerks; and Logistics Clerks. However, some occupations that did not experience a large increase in commencements due to the BAC program also had high cancellation rates, for example Truck Drivers and Storepersons.

\FloatBarrier
\section{Discussion}

\subsection{High-level findings}
Overall, the BAC program had a large positive effect on apprenticeship commencements. There were about 191,000 (between 160,000-220,000) additional commencements during the 7 calendar quarters that the BAC program was open to entrants compared to what modelling suggests would have been expected without BAC. This represents an increase of about 70 per cent on the level of commencements that would likely have occurred in the absence of the BAC. 

The increase in commencements was especially pronounced in non-trade occupations (114,000) but was also observed in trade occupations (77,000). Additional commencements were higher in the first and last quarters of the BAC program. These patterns were broadly consistent across states. It is worth noting that these non-trade occupations were largely not in the priority occupations list and that the new commencements decreased the proportion of commencements in priority occupations \citep{australian_centre_for_evaluation_rapid_2025}.

Cancellation rates were higher for non-trade commencements (7\% increase) during BAC, but slightly lower for trade commencements (0.7\% decrease). Cancellation rates were particularly high for occupations that experienced a strong growth in commencements due to the BAC program.

These results show the effect that we might expect in theory --- that the introduction of a subsidy will incentivise higher commencements. The kinds of firms that have lower completions might be more highly incentivised to take on apprentices in response to this incentive. In addition, we see this concentrated in non-trade occupations because these are more sensitive to incentives having higher effective subsidies \citep{stanwick_issues_2021}. This is especially true as trade apprenticeships tend to take much longer than non-trade to complete but the apprenticeships are most generously subsidised in the first year. For example, a two-year, non-trade apprenticeship will attract the 50\% wage subsidy of the BAC for half of the apprenticeship while a trade apprentice who may be training for four years enjoys that subsidy for only a quarter of their time as an apprentice. If we are to believe the distinction made by \citet{pfeifer_firms_2016} between investment and productive models, it is also not a surprise that non-trade occupations were particularly keen to take on apprentices while the subsidy was high, and then have these apprentices stop their training when the subsidy fell after 12 months.

Indeed, this idea is supported by interviews. Some stakeholders reported there was some amount of sharp practice in non-trade occupations where businesses converted existing workers to apprenticeships for the purpose of receiving the wage subsidy rather than for the purpose of training. This was likely one of the key drivers of the higher cancellation rates for non-trade occupations as employees who did not intend to complete their apprenticeships cancelled their training once their employers had received the generous BAC wage subsidy. While apprenticeships can be very useful for existing employees, and it can be problematic to shut out existing employees from incentives altogether \citep{mcdowell_shared_2011}, it is not surprising that some minority of employers pursuing a productive model of apprentice employment should use these programs in a way that is not aligned with the policy intention. 

It is possible that the high rate of commencements in non-trade apprenticeships in New South Wales and Victoria in the first quarter of the program might indicate sharp practice was particularly prevalent in areas hit hard by the pandemic. It might be that employers shifted existing employees to BAC / CAC apprenticeships as a kind of de facto alternative to the JobKeeper program that subsidised employee wages in 2020 before tapering out at the end of the year and start of 2021 \citep{edwards_variation_2022}. However, interestingly, the cancellation rate for apprentices in these states are lower than the national average so if there was higher rates of sharp practice, the existing employees that were shifted to an apprenticeship were at least completing them.

\subsection{Cost effectiveness}
These findings need to be considered in the context of the program cost. The BAC and CAC programs substantially increased the cost to government of an additional apprenticeship commencement and completion. Additional apprenticeships are those above the counterfactual level of apprenticeship commencements and completions that would be expected without the BAC and CAC in place. This cost effectiveness modelling can be found in \citet{australian_centre_for_evaluation_rapid_2025}. 

The two programs combined led to a cost per commencement of approximately \$40,000 and a cost per additional completion of approximately \$80,000. This led to a cost to government per apprentice that was five times higher than in the pre-COVID years \citep{australian_centre_for_evaluation_rapid_2025}. Government would not expect cost-per-apprentice to remain the same, after all, BAC and CAC represented a significant investment during an extraordinary time, but it raises the question of whether this additional cost is worth these outcomes. This paper is not a full economic evaluation and robust quantitative exploration of this question is outside its scope. For example, with additional data in a longer time series, a more complete economic evaluation could assess the change in earnings for apprentices (and consequential impacts on government revenue and social security expenditure). However, this cost should be considered in light of the mixed results of the program and in particular the fact non-trade occupations were more sensitive to the BAC / CAC than trades. These were occupations with lower completion rates, less employment in priority occupations, where an apprenticeship is not usually required for employment, and where there was evidence of sharp practice in use of the program. While non-trade occupations represented a lower cost per commencement and completion than trade occupations, the benefits of non-trade commencements and completions seem to have been lower too.

In addition, looking beyond just the trade / non-trade dichotomy, it is worth reflecting on the kinds of employers that might be attracted into a scheme by high incentives. Per \citet{dickie_fair_2011, bednarz_understanding_2014}, we might expect the kinds of firms to take advantage of these schemes even within trade occupations to be firms providing lower-value apprenticeships from the perspective of government --- i.e. apprenticeships that are unlikely to result in completions. On one level, there is no evidence of this in the sense that cancellation rates fell very slightly among trade apprentices. On another level, the government did boost the financial incentives to retain apprentices with the CAC, and yet there was no real change in retention. It may also be that during a time of hardship, a much wider variety of firms were more sensitive to incentives than might be in normal times.

Comparing these results to Deloitte Access Economics's \citet{deloitte_access_economics_econometric_2012} study of cost effectiveness of other programs in Australia shows that the BAC and CAC are relatively expensive compared to most other programs. That report showed that the cost per additional commencement of the four subsidy programs they studied which targeted commencements and had statistically significant results ranged from \$6,501 for the Apprenticeship Kickstart Bonus, to \$27,747 for the Apprentice Kickstart Extension, \$27,991 for the Support for Adult Australian Apprentices program, and \$47,421 for the Support for Mid Career Apprentices Program. This makes the BAC / CAC a relatively expensive program, but not an outlier. On the cost per completion, the Deloitte Access Economics study found that of the three programs targeting completions which had statistically significant results ranged from \$16,270 per completion for the Innovation Incentive, \$28,460 for the Living Away From Home Allowance, and a substantial outlier in the Commonwealth Trade Scholarship which cost \$529,703 per completion. This makes the \$80,000 figure for the BAC / CAC large, but again not the largest cost per completion. It is worth noting that because the report only estimated cost-effectiveness for estimates that were statistically significant, those programs with the highest cost-per-completion did not have estimates made.

\subsection{Implications for program design}
While the quantitative data sheds some light on the performance of the BAC / CAC, qualitative data collection helped to understand more deeply how the program functioned in practice and shape lessons for similar programs in the future. The BAC program was implemented under tight timeframes during a time of distinct economic uncertainty, and so many of these lessons would not have been apparent at the time the BAC program was first designed and implemented. This provides a particularly valuable opportunity for future policymakers to learn from the BAC experience.

\subsubsection{Growing pains from a rapidly scaled policy}
According to interview participants substantial increases in funding for apprenticeships provided through the BAC program created incentives for ‘sharp practice’. This was particularly evident for existing workers in non-trades occupations. That is, some firms transferred existing workers onto an apprenticeship for the purpose of receiving the wage subsidy, rather than to pursue training. The eligibility criteria for existing workers were tightened within the first few weeks of the BAC program’s launch. The change seems to have reduced the prevalence of sharp practice, though stakeholders interviewed disagree on just how much this helped. Some believed these changes ended sharp-practice entirely, some thought it helped somewhat, and some believed it only had a small effect. In future, this kind of hiring of existing employees should be addressed in the eligibility criteria (although again, per \citet{mcdowell_shared_2011}, a blanket ban on existing employees taking part in apprenticeships is likely not the right answer, as apprenticeships can build human capital for existing employees and their existing knowledge of the employer can make them very valuable compared to newly hired apprentices).

To some extent this sharp practice was due just to the rapid increase in the amount of funding for apprenticeship incentives. Annual government payments to employers for apprenticeships increased approximately ten-fold from pre-COVID to the height of the BAC program (\$360 million for FY2018-19 to \$3.8 billion in FY2021-22) \citep{australian_centre_for_evaluation_rapid_2025}. Had the subsidy been smaller, there would have been less incentive for sharp practice. Similarly, there would have been less risk that service providers, regulators and training providers would struggle to accommodate the surge in demand. That said, it is important not to just focus on this troubling minority of cases. After all, the program did substantially boost commencements and the majority of these apprentices were engaged in line with the policy intent. It may be that some amount of sharp practice was an almost unavoidable price to pay for having such a substantial impact so quickly. Exactly how much sharp practice could have been avoided is difficult to know. Some stakeholders mentioned that a proper consultation process may have avoided some issues but this would have been challenging given the urgency of the program. Instead, the program was rolled out without much consultation and then the design was changed as problems emerged.

Future apprenticeship incentive programs should consider these potential ‘growing pains’ and adapt the program design and implementation accordingly. For example, if possible, it may help if the incentive program starts as a smaller or more targeted program, and then is expanded over time. This could be achieved by initially restricting the program for the highest priority occupations and then later expanding to eligibility. Alternatively, it could mean introducing a smaller subsidy initially, and then gradually increasing the subsidy over time. Finally, it is important to make sure that the other supports needed for successful apprenticeships like quality training places, and other support networks for apprentices are considered as well when trying to increase the number of apprentices through wage subsidies \citep{stanwick_issues_2021, laundy_apprenticeships_2016}.

\subsubsection{Timing of payments}

Stakeholders noted that front-loading of incentives is important to employers as apprentices are least productive to their employers early on, meaning that help is most needed early on in an apprenticeship. The majority of BAC / CAC funding was concentrated in the first year and this clearly boosted commencements to levels above those in previous years even though the counterfactual estimates that employer demand for apprentices would have been lower without the program. By the point support tapers out after BAC, apprentices are presumably productive enough that significant additional support is not needed --- at least if the employers are taking an investment mode approach to apprenticeships \citep{pfeifer_firms_2016} (as opposed to the production mode approach where an apprentice is just a normal subsidised worker whose learning and future value to the business is not prioritised). However, there is of course also a risk that generous front-loading incentivises sharp practice as employers take on existing employees as apprentices for a short period of time in order to get the front-loaded subsidy and then cancel the apprenticeship after that.

In addition, BAC payments were made quarterly which interview subjects argued made the program more attractive than similar incentives before and after this period. For example, between the 2015-16 and 2018-19 financial year, the government's standard support for an electrical apprenticeship was a single payment of \$1500 at six months and another \$2500 at completion. This less regular payment meant it was more difficult to actually support the costs of having an apprentice.

\subsubsection{Rationing of places}
Less positive was the lack of targeting in the program and the corresponding use of a cap (at least initially) to ration BAC participation. While almost all other similar programs target incentives in some way (for example to priority industries or priority geographic areas) the BAC and CAC had no such targeting. This resulted in many apprenticeships in non-priority areas. This is not strictly a problem given the program's erring on the side of broad support to address an economy-wide challenge. However, there are three problems that emerged from this lack of targeting. 

First, the program was initially scoped at 100,000 places and as can be seen in Figure \ref{fig:bac_quarter_effects}, this resulted in a rush of demand. This resulted in a range of problems and perverse incentives. For example, some stakeholders reported businesses signing up more apprentices than they had the capacity to supervise in order to secure places before the cap was reached. 

Second, there were also a range of fairness and administrative concerns around caps. Given that there was a rush for places, the process for an employer to sign up for the programs was time-consuming and employers worried the cap would be reached before they were approved. In addition, TAFEs and registered training organisations were put under strain by the unexpected demand for training \citep{australian_centre_for_evaluation_rapid_2025}. A decision was made in March of 2021 to remove the cap prioritising access to the program over adhering to the original budget. The use of a cap for rationing then would have resulted either in rationing of places purely by ability to sign up quickly (if the cap had stayed in place) or a program that expanded beyond its initial scope (if the cap was lifted as happened in practice). Neither would be a desirable outcome. An alternative approach might have been to target the program more narrowly by sector (for example using the national priority occupations) or on geographic areas with particular need. This would have rationed access by a set of criteria around need rather than on who signed up most quickly. This may have reduced perverse behaviour, created a greater sense of fairness and led to more `valuable' apprenticeships. In fact, in practice, the proportion of apprentices working in priority occupations was notably lower during the BAC / CAC period that it was before or after the program \citep{australian_centre_for_evaluation_rapid_2025}.

Third, a lack of targeting may have increased the amount of sharp practice. If the designation of certain industries as `priority' reflects apprentices being particularly valuable in those industries, these priority industries would therefore presumably be more inclined to an investment rather than a production model for apprentices. Apprentices are likely valuable enough that it is not worth engaging in sharp practice. Finally, some occupations (particularly trade occupations) require an apprenticeship to work in the field while others (for example retail) do not. The economic benefit of credentialing people in the former areas then may be greater than the benefit of supporting those in the latter.

\section{Conclusion}

The BAC and CAC programs represented an effort made by the Australian Government to support apprenticeships during a very unusual time for the Australian economy. The programs succeeded in one of their goals, boosting commencements by 70\% and this led to more apprentices on track to complete their apprenticeships at least in absolute terms. However, it did not substantially increase retention of apprentices. In fact, there was an increase in the rate of cancellations driven largely by more cancellations in non-trade occupations.

Just because there was an increase in the rate of cancellations does not necessarily mean there was no positive effect on retention for some apprentices (indeed there seems to be a small, positive effect for some occupations, particularly trades). Instead it seems like this larger cohort of apprentices and their employers may in some ways have been engaged in `sharp practice', beginning an apprenticeship with no intention of completing it merely to attract the incentive --- for example by moving existing employees into apprenticeship arrangements. This sharp practice seems to have been found mostly in non-trade occupations. These occupations do not have long traditions of apprenticeships and generally lack licensing requirements that would mean an apprenticeship is necessary to work or advance in the occupation. 

Some program design decisions could limit similar sharp practice from occurring in the future, for example in targeting support to certain industries that are deemed high-need or which have occupational licensing systems that require an apprenticeship. This is the approach taken by many other apprenticeship supports in Australia, however, understandably the need to roll-out supports quickly in an emergency and perhaps err on the side of generosity to ensure more potential / actual apprentices and their employers who genuinely needed the support of the program could receive it. These are decisions that might make more sense in an emergency situation than outside of one, but that is also a reason why a real-time monitoring and evaluation can be important, because it can give feedback relatively quickly to support decision-making in tight timeframes.

This work has several limitations that are worth highlighting. It relied on a control-on-observables design for causal inference which is not generally considered best practice in policy evaluation \citep{imbens_causal_2015}. It also did not yet have enough data available to actually measure program completions or cancellations later in the program. It may be that there were other dynamics in play for completions that will not be evident until this data becomes available. 

Finally, there may be external validity threats in trying to generalise this work to designing apprenticeship incentives more generally. It is worth remembering that the period covered in this study was a highly unusual one. For much of this period there were unusual policy settings and social conditions that may have interacted with the BAC and CAC to create unusual outcomes. For example, there were stay-at-home orders and other containment policies in place across areas of Australia for much of this period which may affect the ability of businesses to operate and of apprentices to do their job \citep{edwards_variation_2022}. In addition, there were a range of other extraordinary transfer payments during this period such as the boosted rate of JobSeeker for those unemployed and the JobKeeper wage subsidy which may have affected the incentive effect of the BAC / CAC. While there are certainly still lessons to learn here, one should be careful not to naively assume these kinds of causal effects generalise to other contexts when considering policy choices in the future.

\bibliographystyle{agsm}
\setcitestyle{authoryear,open={(},close={)}}
\bibliography{references}

\clearpage\section*{Appendix A - DEWR review of drivers of factors affecting apprenticeship commencements}

DEWR prepared a list of factors that drive apprenticeships commencements in advance of this evaluation which they shared with the evaluation team. It covers both employer-side (Table \ref{tab:employer_side}) and apprentice-side factors (Table \ref{tab:apprentice_side}). This review informed our approach to the selection of controls in our modelling.

\begin{table}[ht]
  \centering
  \caption{Employer‐side factors affecting the supply of apprenticeship places}
  \label{tab:employer_side}
  \begin{tabular}{@{}p{6cm} p{8cm}@{}}
    \toprule
    \textbf{Category} 
      & \textbf{Factor} \\
    \midrule
    \textit{Industry/occupation-specific factors affecting employers} \citep{nelms_factors_2017, deloitte_access_economics_australian_2020, nechvoglod_cost_2009}
      & \begin{itemize}[itemsep=0.3em,leftmargin=*,label=\textbullet]
          \item Industry or occupation growing and positive outlook
          \item Occupation is licensed requiring an apprenticeship
          \item Desire to ‘give back to industry’
          \item Productivity of apprentice compared to apprentice wages deemed good
          \item Perceived benefit of future qualified worker to the business
          \item Historical industry/employer experience with apprentices
        \end{itemize} \\
     \midrule
    \textit{Other factors affecting employers} \citep{deloitte_access_economics_econometric_2012, deloitte_access_economics_apprentice_2021}
      & \begin{itemize}[itemsep=0.3em,leftmargin=*,label=\textbullet]
          \item Incentive for trade apprentices (moderately effective)
          \item Incentive for non-trade apprentices (highly effective)
          \item Incentives timed to coincide with peak apprentice hiring period
          \item Incentive for existing worker apprentices (highly effective in non-trades, potentially subject to sharp practice)
          \item Incentives to engage businesses new to apprenticeships (very low effectiveness)
          \item Incentive for businesses already with apprentices to hire more (highly effective)
        \end{itemize} \\
    \bottomrule
  \end{tabular}
\end{table}

\vspace{1em}

\begin{table}[ht]
  \centering
  \caption{Apprentice‐side factors affecting the demand for apprenticeship places}
  \label{tab:apprentice_side}
  \begin{tabular}{@{}p{6cm} p{8cm}@{}}
    \toprule
    \textbf{Category} 
      & \textbf{Factor} \\
    \midrule
    \textit{Individual-level factors intrinsic to potential apprentices} \citep{nelms_factors_2017, deloitte_access_economics_australian_2020, dickie_fair_2011, powers_factors_2020}
      & \begin{itemize}[itemsep=0.3em,leftmargin=*,label=\textbullet]
          \item Interested in the occupation
          \item Lower academic ability or left school before completing Year 12
          \item From an English-speaking background or First Nations
          \item Likes ‘working with their hands’ or ‘outdoors’
          \item Male, younger
          \item Likes idea of own business / independence (especially for trades)
        \end{itemize} \\
    \addlinespace
    \textit{Environmental factors affecting potential apprentices} \citep{nelms_factors_2017, deloitte_access_economics_apprentice_2021, dickie_fair_2011, powers_factors_2020}
      & \begin{itemize}[itemsep=0.3em,leftmargin=*,label=\textbullet]
          \item Positive general views of apprenticeship occupations
          \item Lower socio-economic status
          \item Went to a government school
          \item Non-metropolitan environment
          \item Trade occupations in higher proportion in local community
          \item Wage disparity compared to other workers deemed a ‘fair trade’
          \item Apprentice incentives (moderately effective)
        \end{itemize} \\
    \bottomrule
  \end{tabular}
\end{table}

\clearpage\section*{Appendix B - Full results table for commencements regression (main effects only)}

\begin{ThreePartTable}
\small
\begin{TableNotes}
\footnotesize
\item Standard errors in parentheses.
\item $^* p<0.05$, $^{**} p<0.01$, $^{***} p<0.001$.
\end{TableNotes}

\begin{longtable}{llccc}
\caption{Regression results for commencements model}
\label{tab:reg_results_grouped} \\
\toprule
Category & Variable & (All industries) & (Trades) & (Non-trades) \\
\midrule
\endfirsthead

\caption[]{Regression results by period and control category (continued)} \\
\toprule
Category & Variable & (All industries) & (Trades) & (Non-trades) \\
\midrule
\endhead

\midrule
\multicolumn{5}{r}{\textit{Continued on next page}} \\
\endfoot

\bottomrule
\insertTableNotes
\endlastfoot

\multirow{4}{*}{\parbox[c]{2.2cm}{\centering Pre-BAC}}
& Pre-BAC: 2019 Q4
  & -0.062 & -0.057 & -0.075 \\
&
  & (0.057) & (0.081) & (0.090) \\

& Pre-BAC: 2020 Q1
  & 0.060 & 0.028 & 0.103 \\
&
  & (0.060) & (0.085) & (0.102) \\

& Pre-BAC: 2020 Q2
  & -0.199** & -0.029 & 0.027 \\
&
  & (0.092) & (0.144) & (0.150) \\

& Pre-BAC: 2020 Q3
  & -0.060 & -0.001 & 0.111 \\
&
  & (0.083) & (0.104) & (0.128) \\

\midrule

\multirow{7}{*}{\parbox[c]{2.2cm}{\centering BAC period}}
& BAC period: 2020 Q4
  & 0.955*** & 0.870*** & 0.747*** \\
&
  & (0.070) & (0.117) & (0.096) \\

& BAC period: 2021 Q1
  & 0.420*** & 0.323*** & 0.376*** \\
&
  & (0.057) & (0.097) & (0.091) \\

& BAC period: 2021 Q2
  & 0.395*** & 0.231** & 0.398*** \\
&
  & (0.058) & (0.105) & (0.086) \\

& BAC period: 2021 Q3
  & 0.390*** & 0.374*** & 0.540*** \\
&
  & (0.056) & (0.084) & (0.102) \\

& BAC period: 2021 Q4
  & 0.443*** & 0.448*** & 0.345** \\
&
  & (0.063) & (0.096) & (0.102) \\

& BAC period: 2022 Q1
  & 0.653*** & 0.686*** & 0.457*** \\
&
  & (0.063) & (0.141) & (0.099) \\

& BAC period: 2022 Q2
  & 0.778*** & 0.816*** & 0.612*** \\
&
  & (0.060) & (0.106) & (0.090) \\

\midrule

\multirow{4}{*}{\parbox[c]{2.2cm}{\centering Post-BAC}}
& Post-BAC: 2022 Q3
  & -0.232*** & -0.095 & -0.154 \\
&
  & (0.063) & (0.082) & (0.097) \\

& Post-BAC: 2022 Q4
  & -0.001 & 0.018 & 0.024 \\
&
  & (0.059) & (0.083) & (0.089) \\

& Post-BAC: 2023 Q1
  & 0.179*** & 0.236*** & 0.076 \\
&
  & (0.063) & (0.087) & (0.109) \\

& Post-BAC: 2023 Q2
  & 0.060 & 0.113 & 0.085 \\
&
  & (0.061) & (0.086) & (0.096) \\

\midrule

\multirow{15}{*}{\parbox[c]{2.2cm}{\centering Controls}}
& Business confidence
  & 0.001 & 0.001 & 0.003 \\
&
  & (0.001) & (0.002) & (0.002) \\

& Occupation vacancy rate
  & 8.676** & -1.427 & 5.026 \\
&
  & (4.098) & (4.830) & (4.305) \\

& Occupation unemployment rate
  & -6.870** & -6.191 & -1.246 \\
&
  & (3.264) & (3.963) & (3.361) \\

& Log occupation hours worked
  & -7.979 & -26.510 & -56.840 \\
&
  & (29.100) & (62.280) & (76.930) \\

& Squared log occupation hours worked
  & 0.186 & 0.734 & 1.638 \\
&
  & (0.758) & (1.662) & (2.167) \\

& Quarter 2
  & -0.372*** & -0.161*** & -0.646*** \\
&
  & (0.021) & (0.036) & (0.035) \\

& Quarter 3
  & -0.399*** & -0.157*** & -0.614*** \\
&
  & (0.022) & (0.042) & (0.035) \\

& Quarter 4
  & -0.466*** & -0.237*** & -0.605*** \\
&
  & (0.029) & (0.055) & (0.042) \\

& Linear time trend
  & 0.024*** & -0.002 & 0.0005 \\
&
  & (0.006) & (0.007) & (0.006) \\

& Post-2012 Q2 trend change
  & -0.036*** & -0.013* & -0.013* \\
&
  & (0.006) & (0.007) & (0.007) \\

& Pre-2010 level shift
  & 0.392*** & 0.513*** & -0.013 \\
&
  & (0.064) & (0.077) & (0.091) \\

& 2010--2012 level shift
  & 0.424*** & 0.545*** & 0.087 \\
&
  & (0.033) & (0.049) & (0.052) \\

& 2012 Q2 temporary spike
  & 0.488*** & 0.490*** & 0.373*** \\
&
  & (0.060) & (0.090) & (0.092) \\

& 2013 Q2 temporary spike
  & 0.295*** & 0.317*** & 0.279*** \\
&
  & (0.056) & (0.079) & (0.083) \\

& 2013 Q3 temporary spike
  & 0.482*** & 0.421*** & 0.452*** \\
&
  & (0.055) & (0.082) & (0.082) \\

\midrule

\multirow{1}{*}{\parbox[c]{2.2cm}{\centering Intercept}}
& Intercept
  & 95.160 & 249.900 & 503.300 \\
&
  & (279.100) & (583.400) & (682.700) \\

\toprule
\midrule
& Residual Std. Error & 0.051 & 0.071 & 0.075 \\
& Degrees of Freedom  & 41 & 41 & 41 \\
& $R^2$               & 0.9895 & 0.9857 & 0.9720 \\
& Adjusted $R^2$      & 0.9817 & 0.9753 & 0.9515 \\
& F Statistic         & 128.3*** & 94.49*** & 47.43*** \\

\end{longtable}
\end{ThreePartTable}

\end{document}